\documentclass[proceedings]{JHEP3}

\PrHEP{PrHEP hep2001}                   
\conference{International Europhysics Conference on HEP} 

\title{Polyakov conjecture and 2+1 dimensional gravity coupled to particles}

\author{Luigi Cantini\\                                    
       Scuola Normale Superiore,Pisa and INFN Sezione di Pisa, Italy\\ 
       E-mail: \email{cantini@df.unipi.it}}

\author{\speaker{Pietro Menotti}                           
        \thanks{Supported in part by MURST.}\\  
        Dipartimento di Fisica della Universit\`a, Pisa and INFN
Sezione di Pisa, Italy\\                                   
        E-mail: \email{menotti@df.unipi.it}}               

\author{Domenico Seminara \\
        Dipartimento di Fisica della Universit\`a, Firenze and INFN
        Sezione di Firenze, Italy \\
        E-Mail: \email{seminara@fi.infn.it}}

\abstract{A proof is given of Polyakov conjecture about the auxiliary
parameters 
of the $SU(1,1)$ Riemann-Hilbert problem for general elliptic
singularities. Such a result is related to the uniformization of the
the sphere punctured by $n$ conical defects. Its relevance to the
hamiltonian structure of 2+1 dimensional gravity in the maximally
slicing gauge is stressed.}

\begin{document}


\noindent {\bf 1.} 
We shall deal with the proof of a conjecture put forward by
Polyakov \cite{conj} about the accessory parameters of the Riemann-
Hilbert problem 
and with the role such a conjecture plays in 2+1 dimensional gravity
coupled to particles on an open universe. Such a theory  when
formulated in the maximally slicing gauge (Dirac gauge)\cite{BCV} shows a deep
connection with the Riemann- Hilbert problem 
which consists in determining a fuchsian differential
equation whose independent solutions transform according to a given
representation of the $SL(2C)$ group when one encircles the
singularities.
In the second order canonical ADM formulation \cite{MS,CMS} of 2+1 dimensional
gravity a variant of the Riemann-Hilbert problem appears which is described 
in Sec. 2.
Such a problem occurs in determining the conformal factor
$e^{2\sigma}$ in the ADM metric
\begin{equation}
ds^2 = -N^2 dt^2 + e^{2\sigma}(dz +N^z dt)(d\bar z + N^{\bar z} dt).
\end{equation}
In fact the hamiltonian constraint in the maximally slicing gauge can
be written as \cite{MS}
\begin{equation}\label{liou}
4\partial_z\partial_{\bar z} \phi= e^\phi +
4\pi\sum_n\delta^2(z-z_n)(-1+\mu_n) + 4\pi \sum_B\delta^2(z-z_B)
\end{equation}
where 
\begin{equation}
\phi = -2\sigma + \ln \left[\frac{1}{2\pi^2} \sum_n
\frac{P_n}{(z-z_n)} \sum_n \frac{\bar P_n}{(\bar z-\bar z_n)}\right]
\end{equation}
being $z_n$ the particle positions, $P_n$ the canonically conjugate
momenta and $4\pi \mu_n$ the particle masses in Planck units. The $z_B$ are
the zeros of $\sum_n P_n/(z-z_n)$ and are known as 
the apparent singularities. In eq.(\ref{liou}) one recognizes the Liouville
equation with point sources.

The particle equations of motion in the relative coordinates can be
written as \cite{CMS}
\begin{equation}\label{dotz1}
\dot z'_n = -\sum_B\frac{\partial\beta_B}{\partial\mu}\frac{\partial
z'_B}{\partial P'_n}
,~~~~\dot P'_n = \frac{\partial\beta_n}{\partial\mu}+
\sum_B\frac{\partial\beta_B}{\partial\mu}\frac{\partial z'_B}{\partial
z'_n}
\end{equation}
where $\beta_B$ are the accessory parameters related to the apparent
singularities. 
One is faced with the problem of proving the hamiltonian nature of
eqs.(\ref{dotz1}) and possibly to give the hamiltonian. As we shall see this
problem has a straightforward solution if one assumes the validity of
Polyakov's conjecture to which now we come. 

\bigskip
\noindent 
{\bf 2.} Polyakov \cite{conj}
put forward the following conjecture on the
accessory parameters 
$\beta_n$ which appear in the solution of the $SU(1,1)$
Riemann-Hilbert problem
\begin{equation}\label{conjecture}
-\frac{1}{2\pi} d S_P = \sum_n\beta_n dz_n + c.c. 
\end{equation}
where $S_P$ is the regularized Liouville action \cite{takh1},
$S_P = \lim_{\epsilon \rightarrow 0} S_\epsilon $
with 
$$
S_\epsilon [\phi] =\frac{i}{2} \int_{X_\epsilon} (\partial_z\phi 
\partial_{\bar z} \phi +\frac{e^\phi}{2}) dz\wedge d\bar z
+\frac{i}{2}\sum_n g_n\oint_{\gamma_n}\phi(\frac{d\bar z}{\bar z -\bar
z_n}- \frac{d z}{ z - z_n})
$$
\begin{equation}\label{Sepsilon}
+\frac{i}{2}g_\infty\oint_{\gamma_\infty}\phi(\frac{d\bar z}{\bar z}- \frac{d
z}{z}) 
-\pi\sum_n g_n^2 \ln\epsilon^2 -\pi g_\infty^2\ln\epsilon^2,~~~~{\rm
where}~~~~dz\wedge d\bar z = -2i dx\wedge dy 
\end{equation}
and $X_\epsilon$ is the disk of radius $1/\epsilon$ in the complex plane from
which disks of radius $\epsilon$ around all singularities have been
removed; $\gamma_n$ are the boundaries of the small disks and
$\gamma_\infty$ is the boundary of the large disk.
In eq.(\ref{conjecture}) $S_P$ has to be computed on the solution of the
inhomogeneous Liouville equation which arises
from the minimization of the action i.e.
\begin{equation}\label{pic}
4 \partial_z \partial_{\bar z}\phi = e^{\phi} +4\pi\sum_n g_n\delta^2(z-z_n)  
\end{equation}
with 
behavior at infinity $\phi= -g_\infty \ln z\bar
z+O(1)$.  
Such a conjecture plays an important role in the quantum Liouville
theory \cite{takh1,ZT1}
and in the ADM formulation of $2+1$ dimensional gravity
\cite{MS,CMS}. 
The conjecture is interesting in itself as it gives
a new meaning to the rather elusive accessory parameters
\cite{yoshida} of the Riemann-Hilbert
problem. In particular it implies that the form 
$
\omega = \sum_n\beta_n dz_n + c.c.
$
is exact.
Zograf and Takhtajan \cite{ZT1} provided a proof of
eq.(\ref{conjecture}) for parabolic singularities. 
In addition they remark that the same
technique can be applied when some of the singularities are elliptic
of finite order. 
On the other hand in $2+1$ gravity one is faced with
general elliptic singularities and here the mapping technique cannot
be 
applied. 
Picard
\cite{picard} proved that eq.(\ref{pic}) for real $\phi$
with asymptotic behavior at infinity 
$
\phi(z) = -g_\infty\ln(z\bar z) + O(1)$, 
$-1<g_n,~~1<g_\infty$ 
and
$\sum_n g_n +g_\infty < 0$ 
admits one and only one solution (see also \cite{troyanov}). 
The interest of such results is that they solve the following variant
of the Riemann-Hilbert problem: at $z_1,\dots z_n$ we are given not
with the monodromies but with the class, characterized by $g_j$, of
the elliptic monodromies with the further request that all such
monodromies belong to the group $SU(1,1)$. The last requirement is
imposed by the fact that the solution of eq.(\ref{pic}) has to be
single  valued. 

From eq.(\ref{pic}) one can easily prove
\cite{poincare}, 
that 
\begin{equation}\label{bgform1}
e^{-\frac{\phi}{2}}=\frac{1}{\sqrt{8}|w_{12}|}[\bar y_2(\bar z)y_2(z)
- \bar y_1(\bar z)y_1(z)]
\end{equation}
being $y_1,y_2$ two independent solutions of 
\begin{equation}\label{fuchsb}
y''+Q(z)y=0~~~~{\rm where}~~~~Q(z) = \sum_n - \frac{g_n(g_n+2)}{4(z-z_n)^2} +\frac{\beta_n}{2(z-z_n)}. 
\end{equation}
$w_{12}$ is the constant wronskian  
and the $\beta_n$ are the accessory parameters \cite{yoshida}.

\bigskip
\noindent
{\bf 3.} 
The result of Picard assures us that given the position of the
singularities $z_n$ and the classes of monodromies characterized by
the real numbers $g_n$ 
there exists a unique fuchsian equation which realizes $SU(1,1)$
monodromies of the prescribed classes. In particular the uniqueness of
the solution of Picard's equation tells us that the accessory
parameters $\beta_i$ are single valued functions of the parameter $z_n$
and $g_n$. We shall examine in this section how such dependence arises
from the viewpoint of the imposition of the $SU(1,1)$ condition on the
monodromies.  
Starting from the singularity in $z_1$
we can consider the canonical pair of solutions around $z_1$, $y^1_1 =
\zeta^{\frac{g_m}{2}+1}A(\zeta)$, $y^1_2 =
\zeta^{-\frac{g_m}{2}}B(\zeta)$ with $A(\zeta) = 1 + O(\zeta)$,
$B(\zeta) = 1 + O(\zeta)$, $\zeta = z -z_1$,   
i.e. those solutions which behave as a single fractional power
multiplied by an analytic function with first coefficient one. 
Let $(y_1, y_2)$ the solutions which
realize $SU(1,1)$ around all singularities. Obviously all conjugations
with any element of $SU(1,1)$ is still an equivalent solution in the
sense that they provide the same conformal factor $\phi$.
The canonical pairs around different singularities are linearly
related i.e. $(y^1_1, y^1_2) = (y^2_1, y^2_2) C_{21}$.
We fix the conjugation class by setting
$
(y_1, y_2) = (y^1_1, y^1_2) K
$
 with $K = {\rm diag}( k, k^{-1})$ being the overall constant irrelevant
in determining $\phi$. Moreover if the solution $(y_1, y_2)$ realizes
$SU(1,1)$ monodromies around all singularities also $(y_1,
y_2)\times{\rm diag}(e^{i\alpha},e^{-i\alpha})$ accomplishes the same purpose
being ${\rm diag}(e^{i\alpha},e^{-i\alpha})$ an element of
$SU(1,1)$. Thus the phase of the number $k$ is irrelevant
and so we can consider it real and positive. 
This choice of the canonical pairs is always possible in our case. 
If $D_{n}$ denote the diagonal monodromy matrices around $z_n$, we
have that the monodromy around $z_1$ is $D_1$ and the one around $z_2$
is
$
M_2 = K^{-1} C_{12} D_2 C_{21} K   
$,
where with $C_{12}$ we have denoted the inverse of the $2\times 2$
matrix $C_{21}$. 
In the case of three singularities (one of them at infinity) 
by using the freedom on $K$ we can reduce $M_2$ to the $SU(1,1)$ form. 
The possibility of such a choice is assured by
Picard's result 
and in this simple case also by the explicit solution in terms of
hypergeometric functions \cite{BCV,MS}.
We come now to a qualitative description of the case of four
singularities. 
We recall that the accessory parameters $\beta_n$ are bound by two
algebraic relations known as Fuchs relations \cite{yoshida}. Thus
after choosing 
$M_1$ of the form $M_1=D_1 K$,  in imposing the $SU(1,1)$ nature of
the remaining monodromies we have at our disposal three real parameters
i.e. $k$, ${\rm Re}~\beta_3$ and ${\rm Im}~\beta_3$. It is sufficient
to impose the $SU(1,1)$ nature of $M_2$ and $M_3$ as the $SU(1,1)$
nature of $M_\infty$ is a consequence of them.  
As the matrices $M_n= K^{-1}C_{1n}D_nC_{n1}K$ satisfy identically
$\det M_n=1$ and ${\rm Tr} M_n = 2\cos\pi g_n$ we need to impose
generically on $M_2$ only two real conditions e.g. ${\rm Re}~b_2={\rm
Re}~c_2$ and ${\rm Im}~b_2=-{\rm Im}~c_2$. The same for $M_3$. Thus is
appears that we need to satisfy four real relations when we can vary
only three real parameters.  The reason why we need only three and not
four is that for any solution of the fuchsian problem the following
relation among the monodromy matrices is identically satisfied
$
D_1 K M_2 M_3 M_\infty=1.
$
 The above reported discussion can be put on rigorous
grounds \cite{CMS2} for any number of singularities by exploiting the
existence and uniqueness of the solution and using some basic results
of the theory of analytic function of several complex variables,
reaching the result that $\beta_n$ are analytic functions of $z_m,\bar
z_m$ in the neighborhood of any point of the complex plane, except for
a finite number of points.  

\bigskip
\noindent
{\bf 4.} Being defined through a limit procedure the action $S_P$ is somewhat
uncomfortable to work with.
It is
however possible, 
introducing a background field to rewrite $S_P$ as a simple integral.
In the global coordinate system $z$ on $C$  one writes
$\phi = \phi_M + \phi_0+\phi_B$ 
where $\phi_B$ is a background conformal factor which is regular and
behaves at infinity  like $\phi_B = -2\ln(z\bar z)+c_B+O(1/|z|)$ while
$\phi_0$ is given by
\begin{equation}\label{alpha}
\phi_0 = \sum_n g_n \ln|z-z_n|^2 -\alpha \phi_B+c_0~~~~{\rm
where}~~~~\alpha =  - (\sum_n g_n  +g_\infty -2)/2.
\end{equation}
Then we have for $\phi_M$
\begin{equation}
\label{eqphiM} 4\partial_z\partial_{\bar z}\phi_M =
e^{\phi_0+\phi_B+\phi_M}+(\alpha-1)4~\partial_z\partial_{\bar z}
\phi_B.  
\end{equation} 
$\phi_M$ is a continuous function on the
Riemann sphere.  
The action which generates the above equation is
\begin{equation}\label{Saction} 
S= 
\int [ \partial_z\phi_M \partial_{\bar z}\phi_M + \frac{e^{\phi}}{2}+
2(\alpha-1)\phi_M\partial_z\partial_{\bar z}\phi_B] \frac{i dz\wedge
d\bar z}{2}. 
\end{equation} 
The integral in eq.(\ref{Saction})
converges absolutely. 
$S$ computed on the solution of eq.(\ref{eqphiM}) is related to the
original Polyakov action $S_P$ also computed on the solution of
eq.(\ref{eqphiM}) by 
$$ S_P = S - (\alpha-1)^2\int
\phi_B\partial_z\partial_{\bar z}\phi_B \frac{idz\wedge d\bar z}{2} +
2\pi (\alpha -1)^2c_B+ 
$$ 
\begin{equation}\label{SPtoS}
+\pi\sum_m\sum_{n\neq m} g_m g_n \ln|z_m-z_n|^2 +4\pi c_0(1-\alpha).
\end{equation} 
Our aim now is to compute the derivative of $S_P$
with respect to $z_n$. Again computing the derivative of the new
action $S$ is not completely trivial because one cannot take directly
the derivative operation under the integral sign. In fact such
unwarranted procedure would give rise to an integrand which is not
absolutely summable. One can device however a subtraction procedure
\cite{CMS2} 
which allows to perform such operations. The rigorous result for the
derivative, 
using 
the equation of motion (\ref{eqphiM}) is \cite{CMS2}
\begin{equation}
\frac{\partial S}{\partial z_m} = \lim_{\epsilon
\rightarrow 0} 
\int_{X_\epsilon}[\partial_z(\frac{\partial \phi_M}{\partial
z_m}\partial_{\bar z} \phi_M) +
\partial_{\bar z}(\frac{\partial \phi_M}{\partial
z_m}\partial_{z} \phi_M) 
+\frac{\partial \phi_0}{\partial
z_m}\frac{e^\phi}{2}]\frac{i dz\wedge d\bar z}{2}
\end{equation}
which can be computed by using eq.(\ref{eqphiM}) 
to obtain
\begin{equation}\label{contour}
\frac{\partial S}{\partial z_m} = - i g_m \lim_{\epsilon \rightarrow 0}
\oint_{\gamma_{\epsilon}} \frac{1}{z-z_m}
\partial_z\left(\phi_M-(\alpha-1)\phi_B\right) d z.
\end{equation}
Using $\phi_M -(\alpha-1)\phi_B = \phi -\sum_n g_n\ln|z-z_n|^2$ and
the expansion of 
$A=1+c_1\zeta+\cdots$ and $B
=1+ c_2\zeta + \cdots$ which are obtained by substituting into the
differential equation (\ref{fuchsb}) to obtain
\begin{equation}
c_1=-\frac{\beta_m}{2(g_m+2)}~~~~{\rm and}~~~~c_2= \frac{\beta_m}{2g_m}
\end{equation}
finally we have
\begin{equation}
\frac{\partial S}{\partial z_m} = 
-2\pi \beta_m - 2\pi\sum_{n,n\neq m} \frac{g_m g_n}{z_m-z_n} 
\end{equation}
equivalent to Polyakov conjecture eq.(\ref{conjecture}) due to the
relation (\ref{SPtoS}) between $S$ and $S_P$.
From eqs.(\ref{dotz1},\ref{conjecture}) we see that the hamiltonian is
given by $H=\frac{1}{2\pi}\frac{\partial S_P}{\partial \mu}$, because
\begin{equation} 
\frac{\partial H}{\partial P'_n} = -\sum_B \frac{\partial \beta_B}{\partial
\mu}\frac{\partial z'_B}{\partial P'_n}~~~~{\rm and}~~~~-\frac{\partial
H}{\partial z'_n} = \frac{\partial \beta_n}{\partial \mu} +\sum_B
\frac{\partial \beta_B}{\partial  
\mu}\frac{\partial z'_B}{\partial z'_n}.  
\end{equation}
\smallskip\smallskip
\noindent
{\bf Acknowledgments}
We are grateful to Mauro Carfora for pointing out to us reference
\cite{troyanov} and for useful discussions.





\begin{thebibliography}{99}
\bibitem{conj} A.M. Polyakov as reported in refs.\cite{takh1,ZT1}.

\bibitem{BCV} A. Bellini, M. Ciafaloni, P. Valtancoli, Physics Lett. B
357 (1995) 532; Nucl. Phys. B 462 (1996) 453; M. Welling,
Class. Quantum Grav. 13 (1996) 653; Nucl. Phys. B 515 (1998) 436.

\bibitem{MS} P. Menotti, D. Seminara, Ann. Phys. 279 (2000) 282;
Nucl. Phys. (Proc. Suppl.) 88 (2000) 132. 


\bibitem{CMS} L. Cantini, P. Menotti, D. Seminara, Class. Quantum
Grav. 18 (2001) 2253.

\bibitem{takh1} L. A. Takhtajan, Mod. Phys. Lett. A11 (1996) 93;
``Topics in quantum geometry of 
Riemann surfaces: two dimensional quantum gravity'',
Proc. Internat. School Phys. Enrico Fermi, 127, IOS, Amsterdam, 1996.

\bibitem{ZT1} P. G. Zograf, L. A. Takhtajan, Math. USSR Sbornik 60
(1988) 143; Math. USSR Sbornik 60 (1988) 297. 

\bibitem{yoshida} M. Yoshida, ``Fuchsian differential equations'',
Fried. Vieweg \& Sohn, Braunschweig (1987); K. Okamoto,
J. Fac. Sci. Tokio Univ. 33 (1986) 575;J. A. Hempel, Bull. London
Math. Soc. 20 (1988) 97.

\bibitem{picard} E. Picard, Compt. Rend. 116 (1893) 1015;
J. Math. Pures Appl. 4 (1893) 273 and (1898) 313;
Bull. Sci. math. XXIV 1 (1900) 196. 

\bibitem{troyanov} M. Troyanov, Trans. Am. Math. Soc. 324 (1991) 793. 

\bibitem{CMS2} L. Cantini, P. Menotti, D. Seminara, Phys. Lett. B
517 (2001) 203 .


\bibitem{poincare} H. Poincar\'e, J. Math. Pures Appl. (5) 4 (1898) 137.

\end{thebibliography}
\end{document}